\documentclass[intlimits,twoside,a4paper]{article}
\usepackage[cp1251]{inputenc}

\usepackage[eqsecnum]{cmpj3}

\usepackage{graphicx}

\usepackage{bm}
\usepackage{epsfig}
\usepackage{xcolor}
\newcommand{\vvec}[1]{\mathbf{#1}}

\issue{2023}{26}{2}{23301}
\doinumber{10.5488/CMP.26.23301}

\title[On the swelling properties of  pom-pom polymers]%
{On the swelling properties of  pom-pom polymers: impact of backbone length}%

\author[K. Haydukivska, V. Blavatska]{K. Haydukivska \orcid{0000-0002-3118-7010} \refaddr{label1,label2}
\thanks{Corresponding author: \email{wja4eslawa@icmp.lviv.ua}.},
 V. Blavatska \orcid{0000-0001-6158-1636} \refaddr{label1,label3}
}
\addresses{
\addr{label1} Institute for Condensed Matter Physics of the National Academy of Sciences of Ukraine,\\
1 Svientsitskii St., 79011 Lviv, Ukraine
\addr{label2} Institute of Physics, University of Silesia, 75 Pu\l{}ku Piechoty 1, 41-500 Chorz{\'o}w, Poland
\addr{label3} Dioscuri Centre for Physics and Chemistry of Bacteria,
Institute of Physical Chemistry, Polish Academy of Sciences, 01-224 Warsaw, Poland	
}

\date{Received September 19, 2022, in final form November 15, 2022}
\begin{document}

\maketitle

\begin{abstract}
The present work continues our previous studies of pom-pom molecule [K. Haydukivska, O. Kalyuzhnyi, \linebreak V. Blavatska, and J. Ilnytskyi, J. Mol. Liq. {\bf 328}, 115456 (2021); Condens. Matter Phys. {\bf 25}, 23302 (2022)]. The molecule consists of a linear backbone with two branching points at both ends, with functionalities $f_1$ and $f_2$. Here, the main attention is concentrated on studying the impact of the central backbone length  on the configurational charactersitics of complex molecule, such as size and shape ratios.
We apply both a direct polymer renormalization scheme based on continuous chain model and the alternative Wei's method to analyze a set of size and shape properties of pom-pom polymers in dilute solution.  The size ratio of a pom-pom and a chain polymer of the same total molecular mass is calculated with an excluded volume interaction taken into account, and estimates for asphericity are found in Gaussian approximation, whereas for the size ratio we found a monotonous dependence of the length of backbone at different functionalities of side arms.  Results for asphericity show a non-trivial behaviour.  

\end{abstract}
\section{Introduction}

In recent decades a number of methods were developed that permit to synthesize complex polymers with a desired number of branching points and their functionalities, the length of individual branches, the presence of loops, etc.\cite{gregory2012,Polymeropoulos2017,England2010,Hirao11}. Such an interest has arisen due to the strong impact of architecture of individual macromolecules on the expected properties of their melts or solvents  \cite{Duro2015,Jabbarzadeh2003,Long2020}.

 The simplest representative of the multiple branching polymer architecture is the so-called  H-polymer, which  was not only synthesised by a number of different strategies but was also thoroughly studied~\cite{Roovers84,McLeish88}. As generalization of this structure, the pom-pom architecture, containing two branching points of functionalities $f_1$ and $f_2$ was synthesised \cite{bayer1994,gu2017,knauss2002,MM2000,Long2020,Zheng2001}.  A number of studies were dedicated to the properties of such macromolecules in melts \cite{McLeish98,Graham01,Ruymbeke07,Chen10}.  It is important to notice that also solutions of multibranched polymers exhibit  significantly different viscoelastic properties in comparison with the case of  molecules of more simple topologies \cite{knauss2002}. On the other hand, branched polymers are used as  viscosity modifiers in the solution, for example in lubricants because it is a well known fact that branched polymers are characterised by a lower intrinsic viscosity than their linear counterparts of the same total molecular mass. Quantitatively, it is described by the shrinking factor which is equal to a ratio between the viscosities of the branched and liner polymers \cite{Khabaz14,knauss2002,Fer2019}. Some experimental data show a decrease of this ratio with an increase of the branching parameter for pom-pom polymers \cite{knauss2002}. The decrease in viscosity is traditionally related to the decrease of the effective polymer size. This relation between intrinsic viscosity and effective size is described by the Flory-Fox equation \cite{Kok1981}.

The properties of polymer melts and dense solutions strongly depend on the conformational characteristics of individual macromolecules. The study of such properties can be much easier conducted while considering polymers in dilute solutions, when the interaction between different molecules is insignificant and thus the properties of a single molecule can be analyzed \cite{Burchard}. In this scenario, in statistical description of molecules one can find a number of properties  which do not depend on any microscopic details of the macromolecules but rather depend  on the so-called global characteristics, such as as space dimension, quality of the solvent and the  topology of macromolecules \cite{desCloiseaux,deGennes}. As typical examples of such universal properties, one consideres, e.g., the so-called size ratio $g$ of mean-squared gyration radii of the complex molecule $\langle R_g^2 \rangle_{{\rm complex}} $ and that of the simplest linear polymer chain
 $\langle R_g^2 \rangle_{{\rm chain}} $ of the same total molecular weight \cite{Zimm49}, describing the effective extention of the shape of complex polymer architecture in solution, as compared with the linear one.  
 More subtle size characteristics, specific to branched polymers,
are the individual backbone and side branches swelling ratios as well as backbone-to-side branches ratio, sudied in detail in our previous works  \cite{Haydukivska21, Haydukivska22}. 
The elongation of the macromolecule may be also characterised  by considering the universal shape characteristics like the asphericity $A_d$. It describes the deviation of the shape from a spherical one being equal to $0$ for a sphere and reaching the value of $1$ for the rod-like state and can be defined as \cite{Aronovitz,Rudnick86}:
\begin{equation}
\langle A_d\rangle=\frac{1}{d(d-1)}\left\langle\frac{{\rm Tr}\, \hat{\bf S}^2 }{({\rm Tr}\, {\bf S})^2} \right\rangle\label{Ad}.
\end{equation}
Here, ${\bf S}$ is the gyration tensor, $\hat{ {\bf S}}= {\bf S}-\overline{\mu}{\bf I}$ with $\overline{\mu}$ being an average eigenvalue and ${\bf I}$ is a unity matrix. 
In our previous studies \cite{Haydukivska21,Haydukivska22}, we analyzed a case of pom-pom polymer, where both the backbone and side arms are of equal length $L$. 
Here, we consider the central backbone of variable length $aL$ (with variable $a$), which is assumed to be longer than the side chains. This structure is expected to have a more elongated shape than a star like structure.

The layout of the paper is as follows.
We start this paper by calculations of the size  and shape characteristics in terms of continuous chain model in section \ref{Cont}, that is followed by application of the Wei's method in section \ref{Wei}. Results received in both methods are compared and discussed in section \ref{RD} before we finish this work with some final remarks in conclusions.

\section{Continuous chain model}
\label{Cont}
\subsection{The model}
\label{M}
Within the frames of continuous chain model \cite{Edwards}, the linear polymer chains are described as trajectories of length $L_i$ that are parameterised by radius vector  $\vvec{r}_i(s)$ with $s$ changing from $0$ to $L_i$. The Hamiltonian of the pom-pom architecture can be thus presented as:
\begin{eqnarray}
&&H = \frac{1}{2}\sum_{i=1}^{F}\,\int_0^{L_i} \rd s\,\left(\frac{\rd\vvec{r_i}(s)}{\rd s}\right)^2+\frac{u}{2}\sum_{i,j=0}^{F}\int_0^{L_i}\rd s'\int_0^{L_j} \rd s''\,\delta(\vvec{r_i}(s')-\vvec{r_j}(s'')),\label{H}
\end{eqnarray}
with the first term representing a chain connectivity, the second one describing an excluded volume interactions with coupling constant $u$, $L_i=L$ for branches of both side stars, and $L_0=aL$ (for backbone), and $F=f_1+f_2$.

Within this model, the polymer topology is introduced in the definition of the partition function according to:
\begin{eqnarray}
&&Z^{{\rm pom-pom}}_{f_1,f_2}=\frac{1}{Z_0^{{\rm pom-pom}}}\int\,D\vvec{r}(s)\prod_{i=1}^{f_1}\prod_{j=1}^{f_2}\,\delta(\vvec{r_i}(0)-\vvec{r_0}(0))
\delta(\vvec{r_j}(0)-\vvec{r_0}(L_0))\,{\rm e}^{-H},
\label{ZZ}
\end{eqnarray}
with $Z_0^{{\rm pom-pom}}$ being a partition function of Gaussian model (corresponding to the absence of the second term with excluded volume interaction in the Hamiltonian (\ref{H})), trajectory $\vvec{r_0}(L_0)$ being the backbone chain and $\delta$-functions stating that there are $f_1$ and $f_2$ chains starting at its end points. For the further analyses we put $f_1=f_2=f$.

Observables, calculated on the basis of continuous chain model, can be presented as functions of dimensionless coupling constants $u_0=(2\piup)^{-d/2}uL^{2-d/2}$. In the limit $L\rightarrow \infty$, this constant also tends to infinity and an observable becomes divergent. In order to remove these divergences, it was proposed to replace this coupling constant with a renormalized one $u_R$ \cite{desCloiseaux},  which in the same limit reaches a fixed value: $
\lim_{L\to\infty} u_{R}(u_0)=u_R^*.
$
For the model under consideration, the values of fixed points are well known, and in the first order of the $\epsilon=4-d$ --- expansion read \cite{desCloiseaux}:
\begin{eqnarray}
&& {\rm {Gaussian}}: u^*_{R}=0,\quad {\text{at}}\quad d\geqslant4,\label{FPG}\\
&& { \rm {Pure}}:\,\,\,\,\,\,\,\,\,\, u^*_{R}=\frac{\epsilon}{8},\quad {\text{at}}\quad d<4\label{FPP},
 \end{eqnarray}
where (\ref{FPG}) describes an idealized Gaussian model, and (\ref{FPP}) is a coupling constant for the model with excluded volume interaction.
 
\subsection{Universal characteristics: size ratio}

\begin{figure}[t!]
\begin{center}
\includegraphics[width=73mm]{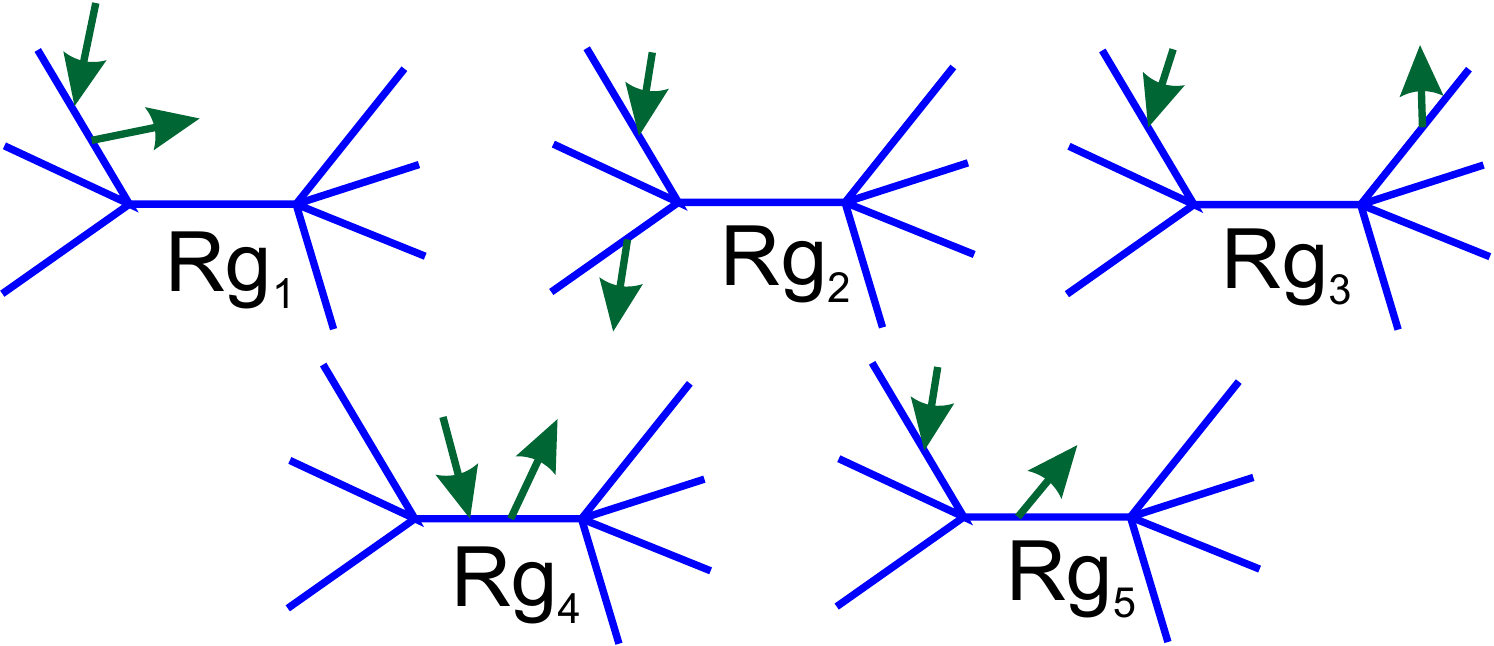}
\caption{ \label{fig:3}(Colour online) Diagrammatic presentations of contributions into $\xi(\vvec{k})$ in Gaussian approximation. The solid lines represent polymer paths  and arrows represent the so-called restriction points $s_1$ and $s_2$.}
\end{center}
\end{figure}

We start our discussion by considering the size ratio of the pom-pom structure and a linear chain of the same molecular mass:  
\begin{eqnarray}
&&g_{c}=\frac{\langle R^2_{g,\text{pom-pom}}\rangle}{\langle R^2_{g,\text{chain}}\rangle}\label{gc},
\end{eqnarray}
with gyration radius defined in terms of continuous chain model as:
\begin{eqnarray}
&&{\langle {R^2_{g}}\rangle} = \frac{1}{2(LF+La)^2}\sum_{i,j=0}^{F}\int_0^{L_j}\int_0^{L_i} \rd s_1\,\rd s_2 \langle(\vvec{r}_i(s_2)-\vvec{r}_j(s_1))^2\rangle.
\end{eqnarray}
We are making use of an identity:
\begin{eqnarray}
&&\langle(\vvec{r}_i(s_2)-\vvec{r}_j(s_1))^2\rangle = - 2 \frac{\rd}{\rd|\vvec{k}|^2}\xi(\vvec{k})_{\vvec{k}=0},\nonumber\\
&&\xi(\vvec{k})\equiv\langle{\rm e}^{-\ri\vvec{k}(\vvec{r}_i(s_2)-\vvec{r}_j(s_1))}\rangle.
\end{eqnarray}
To calculate $\xi(\vvec{k})$ within the path integration approach, we use a diagrammatic technique for which the diagrams for the Gaussian approximation are given in
figures~\ref{fig:3}. These diagrams should be accounted for with some prefactors, as was discussed in our previous works \cite{Haydukivska21,Haydukivska22}. The Gaussian approximation for the pom-pom structure will thus read:
\begin{eqnarray}
&&\langle R^2_{g,\text{pom-pom}}\rangle_0 =\frac{dL}{6(2f+a)^2}\left[f^2(2+a)+f\left(a^2+a-\frac{2}{3}\right)+\frac{a^3}{6}\right].
\end{eqnarray}
Note that the result depends on the relative length of the backbone and side chains. 

Taking into account the excluded volume interaction governed by coupling constant $u_0$, we develop the first order of perturbation theory in the form:
\begin{eqnarray}
&&\langle R^2_{g}\rangle=\langle R^2_{g}\rangle_0\left[1-u_0 A(f_1,f_2,a,d)\right],
\end{eqnarray}
with the expressions for $A(f_1,f_2,a,d)$ provided in the appendix. 

It is a well known practice in the renormalization group approaches to evaluate the  $\epsilon$-expansions for the observable of interest. In order to get a reliable result on their basis, which can be compared with the data of computer simulations or experiments, we need to consider at least the second order of perturbation theory, which is usually a tricky task from mathematical point of view. One of the possibilities is to make use of a Douglas-Freed (DF) approximation (details are provided in the appendix) that gives a good quantitative agreement with simulations \cite{Haydukivska21,Haydukivska22} and experiment~\cite{Douglas84}. 

\begin{figure}[!b]
\begin{center}
\includegraphics[width=73mm]{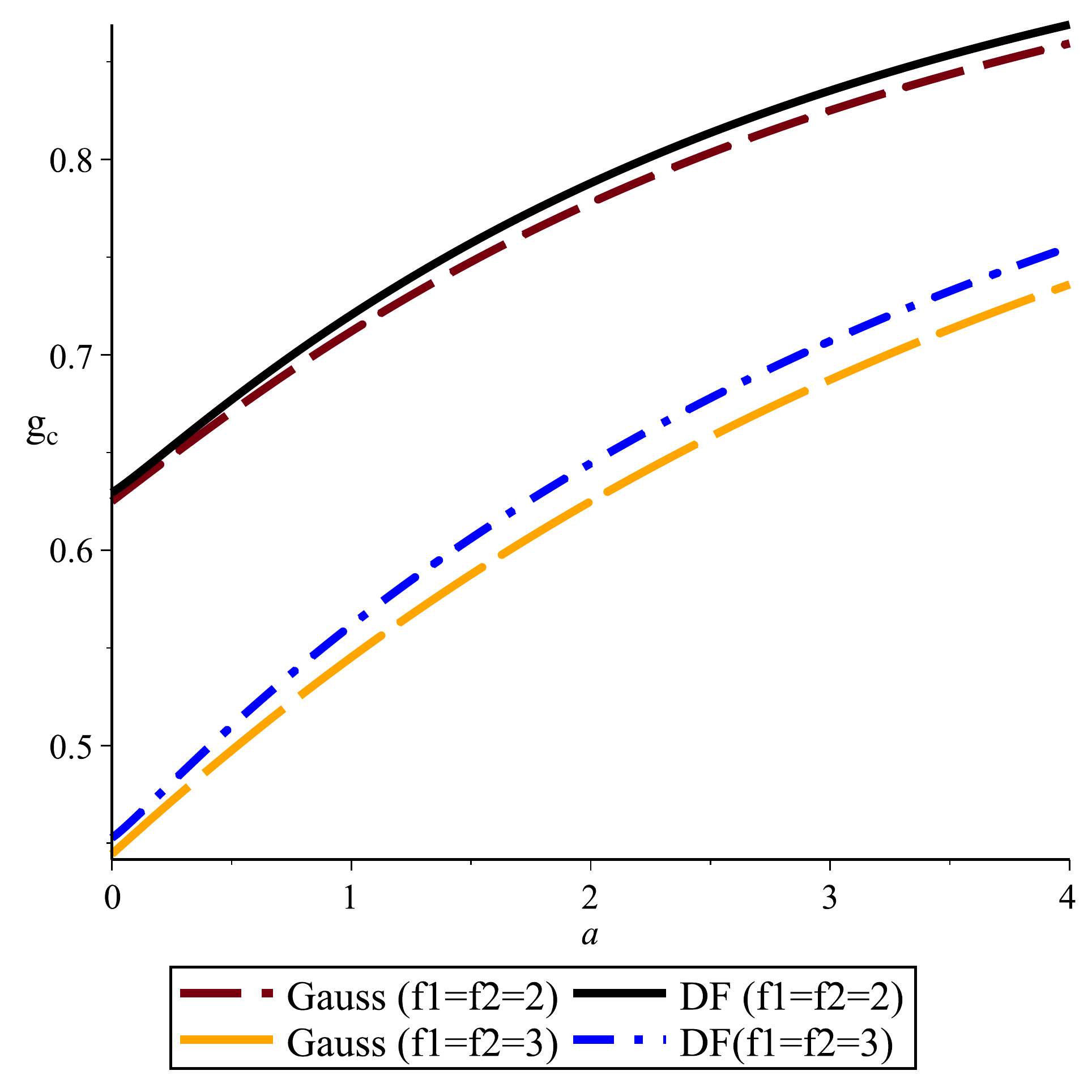}
\caption{ \label{g_c_anal} (Colour online) Size ratio $g_c$ for symmetric case $f_1=f_2=f$ as a function of parameter $a$.}
\end{center}
\end{figure}

To establish the influence of the excluded volume on the size ratio (\ref{gc}), we compare the results of two approximations for a few fixed values of $f$; the data are presented in figure \ref{g_c_anal}. Note that the influence of the excluded volume for a range of values of $a$ is rather small and the general behaviour, tendencies and limits are rather the same. 
In general, let us summarize the main features of the size ratio, as observed in figure \ref{g_c_anal}:
\begin{itemize}
  \item for $a=0$ (backbone of zeroes length), the ratio transforms into a well known result for a star with $2f$ branches versus a linear chain \cite{Blavatska12}.
  \item in the limit $a=\infty$, the ratio tends to $1$. Indeed, in this case the presence of side stars plays practically no role on the behaviour of infinitely long central backbone chain.
  \item in the case of $a=1$ (backbone and side branches of equal lengths), we recover results from our previous work \cite{Haydukivska21}.
\end{itemize}

\subsection{Universal characteristics: asphericity} 

\begin{figure}[b!]
\begin{center}
\includegraphics[width=73mm]{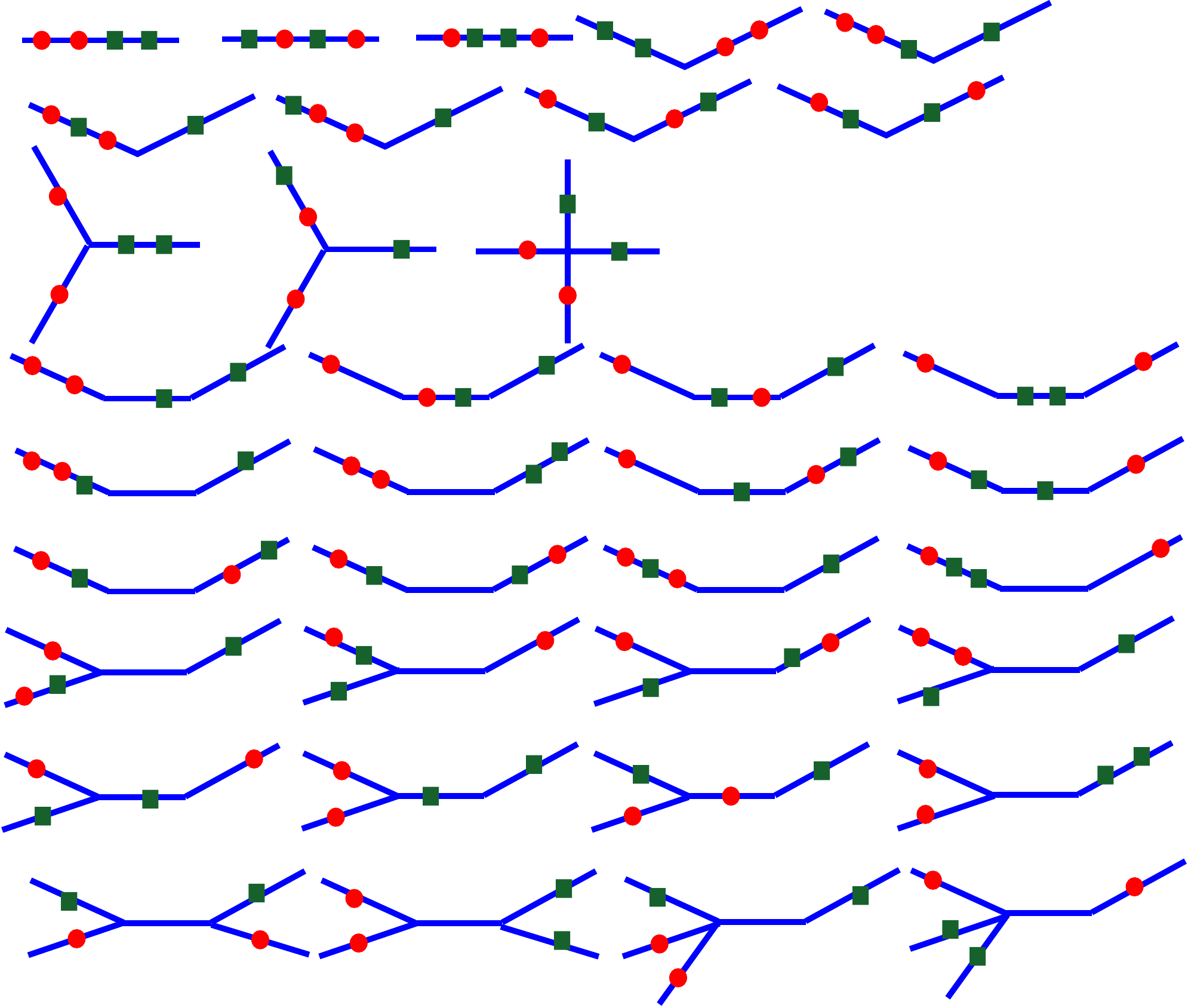}
\caption{ \label{Diagrams_A}(Colour online) Diagrammatic presentation of contributions into $\zeta(\vvec{k}_1,\vvec{k}_2)$ in Gaussian approximation. The solid lines represent the polymer paths and arrows represent the so-called restriction points $s_1,\,s_2,\,s_3$ and $s_4$.}
\end{center}
\end{figure}

We continue our consideration by calculating  the asphericity of the pom-pom structure. An analytical calculation of the asphericity as defined by equation (\ref{Ad}) is rarely possible even in the simple cases. Thus, it was proposed in  \cite{Aronovitz} to consider a slightly different quantity: 
\begin{equation}
\overline{A_d}=\frac{1}{d(d-1)}\frac{\langle{\rm Tr}\, \hat{ {\bf S}}^2 \rangle}{\langle({\rm Tr}\,  {\bf S})^2\rangle} \label{A_d}.
\end{equation}
This can be presented in terms of components of the gyration tensor:
\begin{equation}
\overline{A_d}=\frac{\langle S_{\alpha\alpha}S_{\alpha\alpha}\rangle+d\langle S_{\alpha\beta}\rangle-\langle S_{\alpha\alpha}S_{\beta\beta}\rangle}{\langle S_{\alpha\alpha}S_{\alpha\alpha}\rangle+(d-1)\langle S_{\alpha\alpha}S_{\beta\beta}\rangle}\label{SAD},
\end{equation}
with $S_{\alpha\beta}$ in terms of the continuous chain model given by:
\begin{eqnarray}
&& S_{\alpha\beta}=\frac{1}{2(LF+aL)^2}\sum_{i,j=0}^F\int_0^{L_j}\int_0^{L_i}  (r^{\alpha}_i(s_2)-r^{\alpha}_j(s_1))(r^{\beta}_i(s_2)-r^{\beta}_j(s_1))\rd s_1\,\rd s_2.
\end{eqnarray}
In order to calculate the contributions into (\ref{SAD}), we introduce an identity similarly to the one utilized for the case of gyration radius:
\begin{eqnarray}
&&  r^{\alpha}_i(s_2)-r^{\alpha}_j(s_1))(r^{\beta}_i(s_2)-r^{\beta}_j(s_1))(r^{\alpha}_l(s_4)-r^{\alpha}_m(s_3))(r^{\beta}_m(s_4)-r^{\beta}_j(s_3))=\nonumber\\
&&\frac{\rd}{\rd k^{\alpha}_1}\frac{\rd}{\rd k^{\beta}_1}\frac{\rd}{\rd k^{\alpha}_2}\frac{\rd}{\rd k^{\beta}_2}\zeta(\vvec{k}_1,\vvec{k}_2)|_{\vvec{k}_1=\vvec{k}_2=0}
\end{eqnarray}
with $\zeta(\vvec{k}_1,\vvec{k}_2)={\rm e}^{-\ri\vvec{k}_1(\vvec{r}_i(s_2)-\vvec{r}_j(s_1))}{\rm e}^{-\ri\vvec{k}_2(\vvec{r}_l(s_4)-\vvec{r}_m(s_3))}$. The contributions into this function are calculated using the diagrams presented in figure \ref{Diagrams_A} for a Gaussian case.

\begin{figure}[b!]
\begin{center}
\includegraphics[width=73mm]{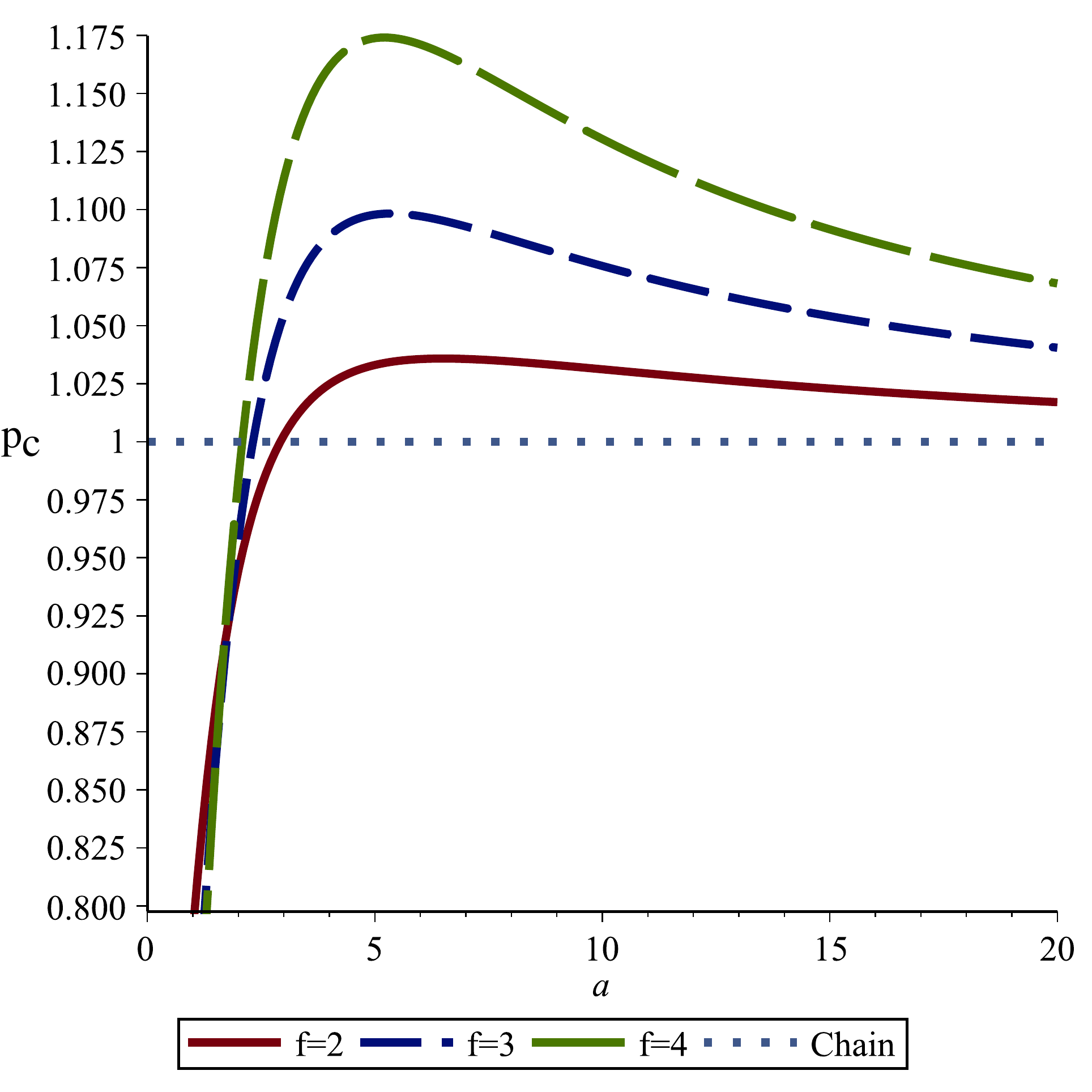}
\caption{ \label{A_CCM_graph}(Colour online) Asphericity ratio $p_c$ for symmetric case $f_1=f_2=f$ as function of the parameter $a$.}
\end{center}
\end{figure}

Performing the calculations we receive the expression:
\begin{equation}
\overline{A_d}=\frac{2(2+d)C_1(f_1,f_2,a)}{5 d C_2(f_1,f_2,a)+4C_1(f_1,f_2,a)} , \label{A_CCM} 
\end{equation}
with $C_1(f_1,f_2,a)$ and $C_2(f_1,f_2,a)$ given by:
\begin{eqnarray}
&&C_1(f_1,f_2,a)=f_2^2(15f_2-14)+f_1^2(15f_1-14)+f_1a(6a^4+15a^3f_1+20a^2+15a+30f_1-24)\nonumber\\
&&+a^6+f_2a(6a^4+15a^3f_2+20a^2+15a+30f_2-24)+f_1f_2(45f_2+45f_1-28\nonumber\\
&&+30a(2f_1+2f_2+2+(6f_1f_2-3f_1-3f_2+7)a+(f_1+f_2+2)a^2+a^3)), \label{A_C1}\\
&&C_2(f_1,f_2,a)=f_2^2(3f_2-2)^2+f_1^2(3f_1-2)^2+36f_1^2f_2(f_1-1)+6f_1^2(3f_1-2)a+3f_1^2(6f_1-1)a^2\nonumber\\
&&+4f_1(6f_1-1)a^3+3f_1(3f_1+2)a^4+6a^5f_1+36f_1f_2^2(f_2-1)+6f_2^2(3f_2-2)a+3f_2^2(6f_2-1)a^2\nonumber\\
&&+4f_2(6f_2-1)a^3+3f_2(3f_2+2)a^4+6a^5f_2+a^6+2f_1f_2(27f_1f_2+4+(18(f_1+f_2)^2+6f_1\nonumber\\
&&+6f_2+6)a+(18f_1f_2+27f_1+27f_2+33)a^2+(9f_1+9f_2+42)a^3+15a^4)\label{A_C2} .
\end{eqnarray}
Again, at $f_1=f_2=1$ and any values of $a$, we recover the result for the linear chain \cite{Rudnick86}, whereas for $f_2=0,f_1=f,a=0$, the asphericity of a single star is recovered \cite{Blavatska12}. Following the same idea as with size ratio in previous subsection, let us introduce the asphericity ratio
\begin{equation}
p_c=\frac{\overline{A_d}_{\text{pom-pom}}}{\overline{A_d}_{\text{chain}}},
\end{equation}
which is usufel in comparing the shape properties of complex polymer and the corresponding linear chain molecule.

Since in the present work we are interested in the influence of the relative length of the backbone, we provide some results for fixed values of branching parameters in figure \ref{A_CCM_graph}.
Since within the continuous chain model we are restricted to the calculation of (\ref{A_d}) rather than (\ref{Ad}), we can receive only a qualitative description with this approach. And since the behaviour of the Gaussian model and the model with excluded volume are rather similar (see figure \ref{g_c_anal}), we restrict ourselves with Gaussian case calculation for the asphericity. An additional bonus to this is a possibility to compare the different averaging with the Wei's method described in the next section.

\section{Wei method and analytical approach to eigenvalue problem of Kirchhoff matrix}
\label{Wei}
Any complex polymer structure 
can be described as a mathematical graph (network), where the individual monomers are presented as vertices, and the chemical bonds between monomers as links between them. The chemical functionalities of monomers are then equal to the degrees of corresponding vertices.
The Wei's method \cite{Wei} is applicable in evaluating the size and shape properties of polymer network of any topology, if  the
Kirchhoff matrix and its eigenvalues are defined. 
For the polymer structure of total number of $M$ monomers, Kirchhoff $M\times M$ matrix ${\bf K}$ is defined as follows. 
Its diagonal elements $K_{ii}$ equal
the degree of vertex $i$, whereas the   non-diagonal elements $K_{ij}$ equal  $-1$ when the vertices $i$ and $j$ are adjacent and $0$ otherwise. 

 Let $\lambda_2,\ldots,\lambda_M$ be $(M-1)$ non-zero eigenvalues of the $M\times M$
Kirchhoff matrix 
\begin{equation}
    {\bf K}{\bf Q}_i =\lambda_i {\bf Q}_i,\,\, \,\,\,\,\,i=1\ldots M 
    \end{equation}
 ($\lambda_1$ is always $0$). 
The
asphericity  in $d$ dimensions is thus given by \cite{Wei,Ferber15}:
\begin{equation}
\langle A_d \rangle =\frac{d(d+2)}{2}\int_0^{\infty} {\rm d} y \sum_{j=2}^{M}\frac{y^3}{(\lambda_j+y^2)^2}\left[ \prod_{k=2}^{M} 
\frac{\lambda_k}{\lambda_k+y^2}\right ]^{d/2}.
\label{awei}\end{equation}

To evaluate the expressions for the set of eigenvalues of the pom-pom structure, we  follow the scheme developed in the original work by Zimm and Kilb \cite{Zimm59} for the case of star-like polymers.
We represent the components of eigenvectors by continuous eigenfunctions.  
Let us introduce notations $Ql_i(s)$, $Qr_i(s)$ with $s=0,\ldots,L$,  $i=1,\ldots, f$ for eigenfunctions corresponding to the``left-hand'' and ``right-hand''  stars, and $Qc(s)$ with $s=0,\ldots,aL$ for the central backbone  chain.  The total number of eigenvalues is thus given by $M=(2f+a)L$.
Taking into account the structure of Kirchhoff matrix for considered structure,  
we may write the eigenvalue equations for the internal points of each branch of side stars in the form:
\begin{eqnarray*}
&&{\bf K}Ql_i(s)=-(2Ql_i(s)-Ql_i^{s-\delta}-Ql_i^{s+\delta}),\\
&&{\bf K}Ql_r(s)=-(2Qr_i(s)-Qr_i^{s-\delta}-Qr_i^{s+\delta}),
\end{eqnarray*}
for the end points:
\begin{eqnarray*}
&&{\bf K}Ql_i(L)=-(Ql_i(L)-Ql_i(L-\delta)),\\
&&{\bf K}Qr_i(L)=-(Qr_i(L)-Qr_i(L-\delta)),\\
\end{eqnarray*}
and for the {central branching points} of two side stars:
\begin{eqnarray*}
{\bf K} Ql_i(0)=-\Bigl( \sum_{j=1}^f Ql_j(0)+Qc(0)-\sum_{j=1}^f Ql_j(\delta)-Qc(\delta)\Bigr) ,\label{center1}\\
{\bf K} Qr_i(0)=-\Bigl( \sum_{j=1}^f Qr_j(0)+Qc(0)-\sum_{j=1}^f Qr_j(\delta)-Qc(\delta)\Bigr) ,\label{center2}
\end{eqnarray*}
where $i=1,\ldots,f$ and $\delta$ is a small parameter. 
The eigenvalue problem is thus reduced to the general equation
\begin{equation}
    -\delta^2\frac{{\rm d}^2 Q^2(s)}{{\rm d}s^2}=\lambda_iQ(s),
\end{equation}
and the solution can be presented in the form
\begin{equation}
Q_n(s)=A_1\cos(ks)+A_2\sin(ks),\,\, k=\sqrt{\lambda}.
\end{equation}

\begin{figure}[t!]
\begin{center}
\includegraphics[width=73mm]{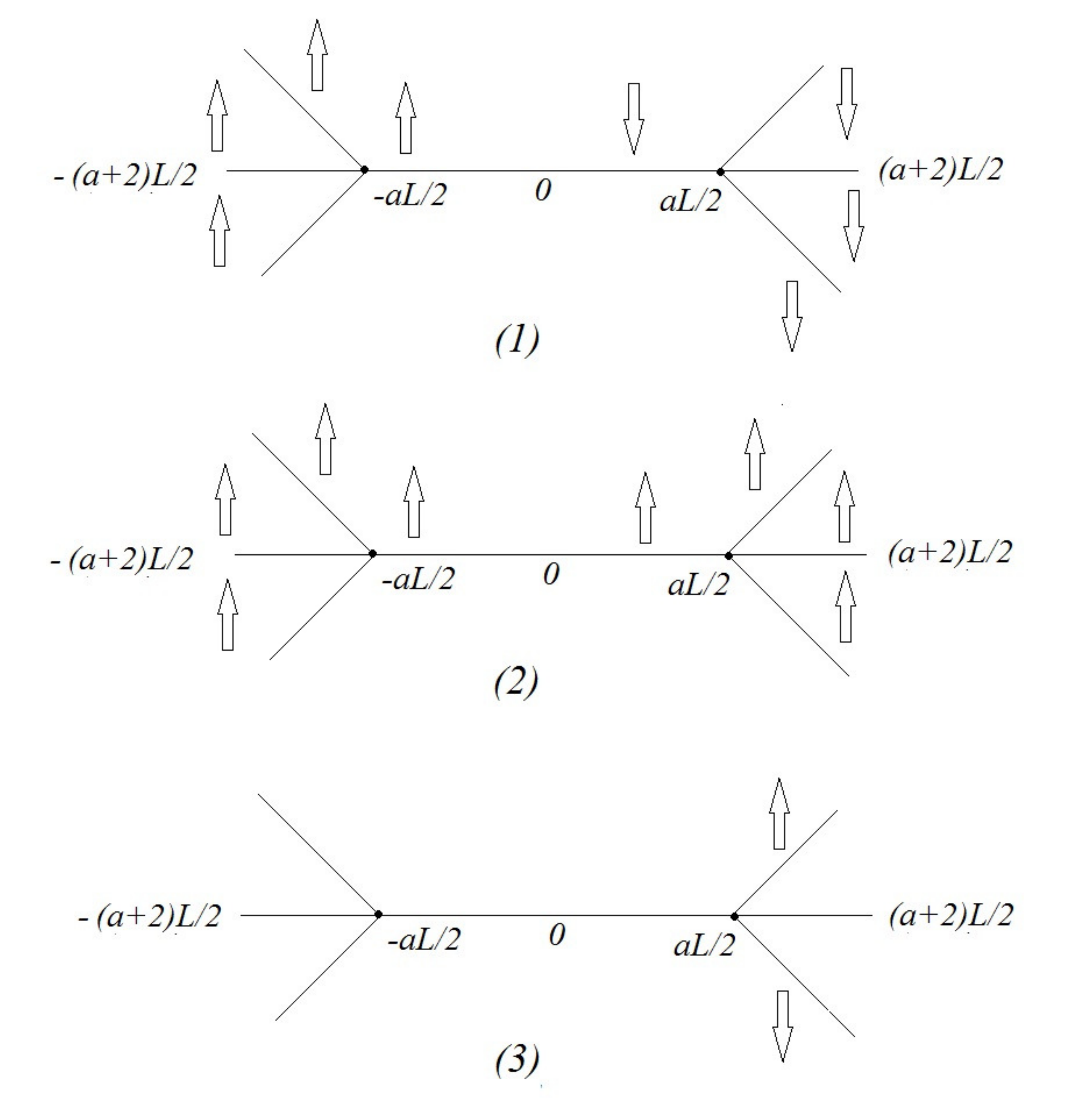}
\caption{ \label{scheme} Schematic presentation of possible vibration modes in pom-pom molecule:  asymmetric (1), symmetric (2), independent asymmetric modes of side stars (3).}
\end{center}
\end{figure}

For convenience of the following evaluation, let us consider the middle of the central backbone as reference point, so that the branching points have coordinates $aL/2$ and $-aL/2$, and the end points $(a+2)L/2$ and $-(a+2)L/2 $, correspondingly (see figure \ref{scheme}). 
The boundary conditions are thus imposed as: 
\begin{enumerate}
    \item 
    $\frac{{\rm d}Qr_i(s)}{{\rm d}s}|_{s=(a+2)L/2}=0,\,\,\,\frac{{\rm d}Ql_i(s)}{{\rm d}s}|_{s=-(a+2)L/2}=0$,
    \item $Qc(aL/2){=}Qr_{i}(aL/2),\,\,\, Qc(-aL/2){=}Ql_{i}(-aL/2)$,
    \item  $\sum\limits_{i=1}^fQr_{i}(aL/2){+}Qc(aL/2){=}0$,\,\,\,
    $\sum\limits_{i=1}^fQl_{i}(-aL/2){+}Qc(-aL/2){=}0$.
\end{enumerate}

{\bf{Antisymmetric eigenfunctions}}

Let us start with the antisymmetric solutions with $Qc(s)=-Qc(-s)$, $Qr_{i}((a+2)L/2+s)=-Ql_{i}(-(a+2)L/2-s)$. We use the following ansatz:
\begin{eqnarray}
&& Qc(s)=\sin(ks),\nonumber\\
&& Ql_{i}(s)=-B\cos(k((a+2)L/2+s)),\,i=1,\ldots,f,\nonumber\\
&& Qr_i(s)=B\cos(k((a+2)L/2-s)),\,i=1,\ldots,f.
\end{eqnarray}
Solving the boundary conditions we find
\begin{eqnarray}
&&B\cos(kL)=-\sin(k \,aL/2), \label{eqas1} \\ 
&&-fB \sin(kL)=\cos(k \,aL/2).\label{eqas2}
\end{eqnarray}
These equations are solved by
\begin{eqnarray}
f\tan(kL)\tan(k \,a L/2)=1.\label{asmain}
\end{eqnarray}
In the case when $a=1$, equation (\ref{asmain}) is easily solved giving two branches of solutions:
\begin{eqnarray}
k_i=2/L \arctan \left(\frac{1}{\sqrt{2f-1}}+n\piup \right), \quad i=1,\ldots,L/2,\\
k_i=-2/L \arctan \left(\frac{1}{\sqrt{2f-1}}+n\piup \right), \quad i=0,\ldots,L/2,
\end{eqnarray}
thus resulting in total in $L$ eigenvalues.
Otherwise, there are $a$ branches in the case of even $a$ and
$(a+1)$ branches for odd $a$.

One more solution of equation \ref{eqas2} is obtained from the condition, then the derivatives of eigenfunctions at branching points equal zero, thus leading to:
\begin{eqnarray}
&&\cos(k\,a L/2)=0 \to k=\frac{(2n+1)\piup}{a L},\quad n=0,1,2,3\ldots,\\
&& \sin(kL)=0 \to k=\frac{n\piup}{L}, \quad n=1,2,3\ldots.
\end{eqnarray}
The simultaneous solution of these equations results in
\begin{equation}
k_n=\frac{(2n+1)\piup}{a L},\label{asad}
\end{equation} 
with $n$ values obeying the condition that $(2n+1)/a$ is integer number, giving additional $L/2$ eigenvalues. Note, that the last condition holds only for the cases of odd $a$. There are additional $L/2$ eigenvalues obtained from this condition.  

{\bf{Symmetric eigenfunctions}}

Here, we are looking for the symmetric solutions with $Qc(s)=Qc(-s)$, $Qr_{i}((a+2)L/2+s)=Ql_{i}(-(a+2)L/2-s)$. We use the following ansatz:
\begin{eqnarray}
&& Qc(s)=\cos(ks),\nonumber\\
&& Ql_{i}(s)=B\cos(k((a+2)L/2+s)),\quad i=1,\ldots,f,\nonumber\\
&& Qr_i(s)=B\cos(k((a+2)L/2-s)),\quad i=1,\ldots,f.
\end{eqnarray}
Solving the boundary conditions we find
\begin{eqnarray}
&&B\cos(k((a+2)L/2-a))=\cos(k\,aL/2), \label{eqs1} \\
&&fB \sin(k((a+2)L/2-a))=-\sin(k\,aL/2).\label{eqs2}
\end{eqnarray}
These equations are solved by
\begin{eqnarray}
f\frac{\tan(kL)}{\tan(k\, aL/2)}=-1.\label{ssmain}
\end{eqnarray}
In the case when $a=1$, equation (\ref{ssmain}) is easily solved giving two branches of solutions:
\begin{eqnarray}
k_i=2/L \arctan \left({\sqrt{2f-1}}+n\piup \right), \quad i=1,\ldots,L/2,\\
k_i=-2/L \arctan \left({\sqrt{2f-1}}+n\piup \right), \quad i=0,\ldots,L/2,
\end{eqnarray}
thus giving $L$ eigenvalues.
Otherwise, there are $a$ branches in the case of even $a$ and
$a+1$ branches for odd $a$.

\begin{figure}[t!]
\begin{center}
\includegraphics[width=73mm]{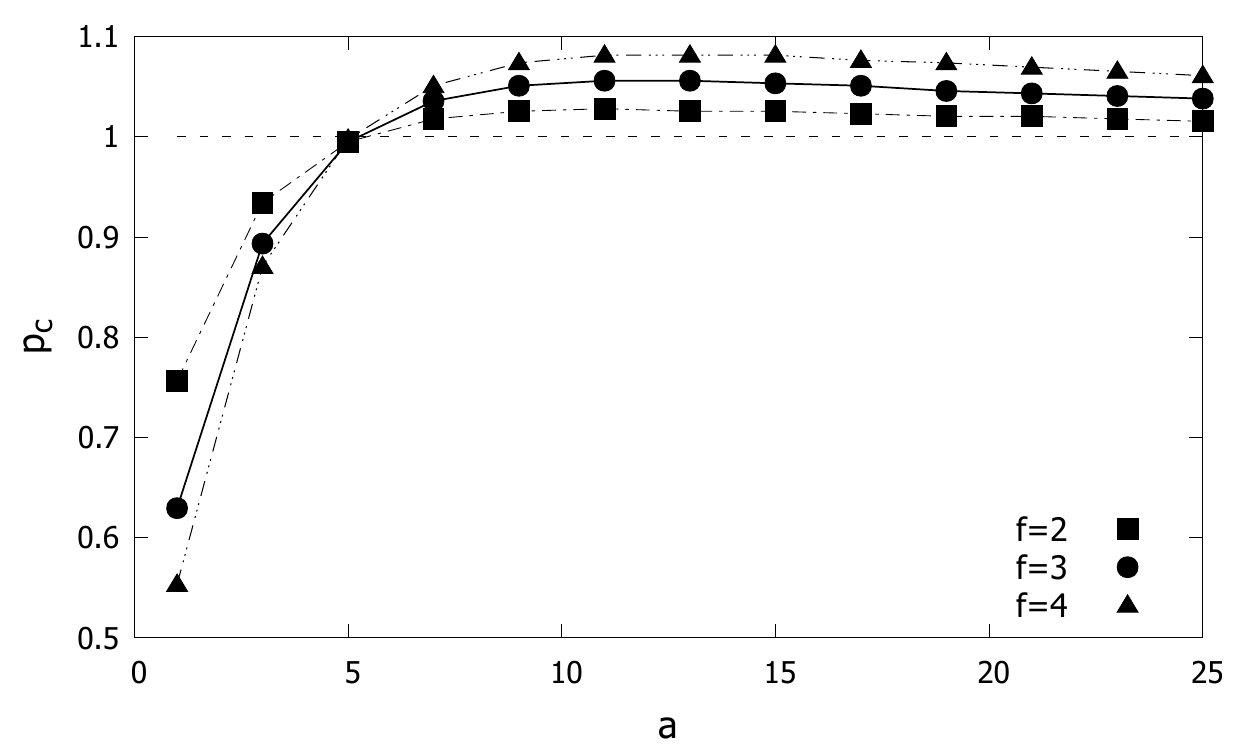}
\caption{ \label{aratio_eigen} Asphericity ratio $p_c$ for symmetric case $f_1=f_2=f$ as function of the parameter $a$ based on on substituting the analytically derived set of eigenvalues in equation (\ref{awei}).}.
\end{center}
\end{figure}

One more solution of equation \ref{eqs2} is obtained from the condition, then the derivatives of eigenfunctions at branching points equal zero, thus leading to:
\begin{eqnarray}
&&\sin(k \, aL/2)=0 \to k=\frac{2n\piup}{a L},\quad n=1,2,3\ldots,\\
&& \sin(kL)=0 \to k=\frac{n\piup}{L},\quad n=1,2,3\ldots.
\end{eqnarray}
The simultaneous solution of these equations results in
\begin{equation}
k_n=\frac{2n\piup}{a L} \label{ssad},
\end{equation}
with $n$ values obeying the condition that $2n/a$ is integer number, this giving additional $L/2$ eigenvalues.

{\bf{Independent asymmetric eigenfunctions of two side stars}}

There are additional functions, having nodes at the branching points of either the side stars, when only two branches of the corresponding star are excited in antisymmetric manner. In this way, the corresponding eigenfunctions coincide with those of star polymer derived in  \cite{Zimm59}.
There are $2(f-1)$ independent eigenfunctions (with degenerate values of $k$) given by: 
\begin{equation}
Q(s)=\sin(k(s-aL/2)), Q(s)=-\sin(k(s-aL/2)).
\end{equation}
They correspond to 
\begin{equation}
k=\frac{(2n+1)\piup}{2L}, n=0,1,2,\ldots, \label{instars}
\end{equation}
giving $2(f-1)L$ eigenvalues.

Thus, the complete set of eigenvalues of Kirchhoff matrix of pom-pom polymer structures is given as $\lambda_i=k_i^2,\,\,i=1,\,\,\ldots,(2f+a)L$, with $(a+1)L/2$ values of $k$ obtained on the base of equation~(\ref{asmain}), $L/2$  of equation (\ref{asad}), $(a+1)L/2$ values of equation (\ref{ssmain}), $L/2$ of equation (\ref{ssad}) and $2(f-1)L$ values given by equation~(\ref{instars}). An estimate for asphericity ratio based on substituting this set of eigenvalues in equation (\ref{awei}) is presented in figure \ref{aratio_eigen}.     

\section{Results and discussion}\label{RD}

The aim of the present study was to describe the impact of the backbone length on the universal properties of the pom-pom polymer. For those purposes, we considered the size and shape properties, such as size ratio and asphericity. 

Our results for the size ratio as defined by equation (\ref{gc}) are received in the framework of the continuous chain model with excluded volume interaction accounted for through the usage of the Douglas-Freed approximation. The results at some fixed values of the branching parameter $f$ are plotted in figure \ref{g_c_anal}. Apart from the very small differences in values between the Gaussain polymers and polymers with excluded volume interactions, it is  interesting to note that the side branches become unimportant rather quickly with increasing the parameter $a$ (the ratio tends to $1$), reflecting the increasing influence of the backbone on the pom-pom characteristic size. 

The asphericity for the case of ideal Gaussain chain is calculated using two different averaging schemes within the frameworks of two different approaches. For the continuous chain model as was mentioned above we used the averaging in as given by (\ref{A_d}) and for the Wei's method by (\ref{Ad}). Due to the different averaging definitions, an absolute value comparison is impossible, thus we consider a ratio:
\begin{equation}
 \alpha(a)=\frac{A^{\text{pom-pom}}_d}{A^{\text{chain}}_d}, \label{adratio} 
\end{equation}
with $A_d$ being ether $\overline{A_d}$ or $\langle A_d \rangle$. This allows us to provide a relative comparison of not only the averaging methods but also the topologies. In both calculation schemes (see figures \ref{A_CCM_graph}, \ref{aratio_eigen}) we can see two distinct regions: at small values of parameter $a$, the ratio is smaller than 1 (the shape of pom-pom structure is more symmetric than 
that of a chain), whereas at $a$ larger than some critical value, the situation is the opposite, while it tends to 1 with $a\to\infty$. 
It is interesting to note that for the large values of $a$, the asphericity of pom-pom structure is only slightly (under 10\%) larger than the corresponding value of the linear chain in both averaging schemes. This indicates a non-trivial influence of the side arms even in the cases where $g_c$ is around $1$. A larger value of asphericity corresponds to the elongation of the macromolecule under the influence of the side arms. Note that the calculations of the shape characteristics were conducted only for the ideal Gaussian chain, that can be treated as an approximation of the so-called $\Theta$-solution \cite{desCloiseaux}, while the behaviour for the polymers in good solutions (with taking into account the excluded volume effect) is expected to be qualitatively similar. 

\section{Conclusions}

This work is a continuation of a cycle of our studies devoted to the analysis of universal conformational properties of pom-pom polymers. Our previous works \cite{Haydukivska21,Haydukivska22} were mainly concentrated on the influence of the branching parameters of side stars on side and shape properties of macromolecule,  whereas here we finally address the question of the impact of the backbone length. Within the frames of the continuous chain model,  we evaluated the estimates for  the size ratio $g_c$ that compares the effective size of the pom-pom topology in a solvent to that of a linear chain with the same molecular mass. We find that as the length of the backbone increases (with an increase of parameter $a$),  the ratio monotonously tends toward $1$, thus for the large value of $a>5$ the characteristic size of the chain and pom-pom is rather similar.

The analysis of the shape characteristics was conducted only for the ideal Gaussian case because in this case it was possible to obtain the exact results for both approaches used: the  continuous chain model framework with application of path integration method and the Wei's method based on evaluation of the set of eigenvalues of Kirchhoff matrix of the graph corresponding to the topology of complex polymer under consideration.  In both cases we find  that the pom-pom molecules are more elongated and asymmetric  as compared to the chain-like topology when the backbone length considerably exceeds the length of side branches. These differences may play an important role in the rheological properties of the polymer solution.

 \section*{ Acknowledgements}

K.H. would like to acknowledge the support from the National
Science Center, Poland (Grant No. 2018/30/E/ST3/00428).
V.B. is grateful for support from the U.S. National Academy of Sciences (NAS) and the Polish Academy of Sciences (PAS) to scientists from Ukraine.
\allowdisplaybreaks
\section*{Appendix}
\setcounter{equation}{0}
\renewcommand{\theequation}{A.\arabic{equation}}

\begin{figure}[b!]
\begin{center}
\includegraphics[width=73mm]{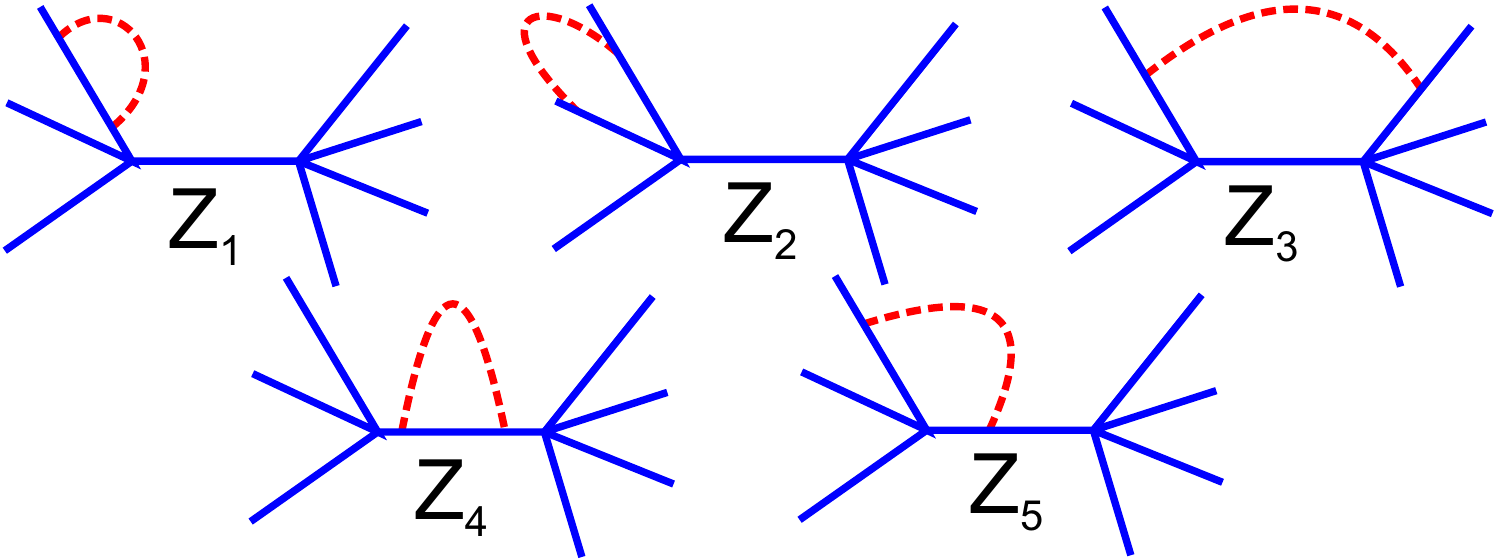}
\caption{ \label{fig:2}(Colour online) Diagrammatic presentations of contributions into partition function up to the first order approximation in coupling constant $u$.
The solid lines are schematic presentations of polymer paths and dash line represents a two monomer excluded volume interaction.}
\end{center}
\end{figure}

Here, we give expressions for contributions into the partition function with taking into account the excluded volume interaction.  Corresponding diagrammatic presentations are given in figure \ref{fig:2}. Diagram $Z_1$ is accounted with prefactor $2f$ times, diagram $Z_2$ comes with $f(f-1)$, diagram $Z_3$ with $f^2$, diagram $Z_5$ with $2f$  $Z_4$ only once. 
The analytical expressions read:
\begin{eqnarray}
&&Z_1=\frac{u(2\piup)^{-d/2}L^{2-d/2}}{(1-d/2)(2-d/2)},\\
&&Z_2=\frac{u(2\piup)^{-d/2}L^{2-d/2}(2^{2-d/2}-2)}{(1-d/2)(2-d/2)},\\
&&Z_3=\frac{u(2\piup)^{-d/2}L^{2-d/2}}{(1-d/2)(2-d/2)}\left[(2L+Lc)^{2-d/2}-2(L+L_c)^{2-d/2}+L_c^{2-d/2}\right],\\
&&Z_4=\frac{u(2\piup)^{-d/2}L_c^{2-d/2}}{(1-d/2)(2-d/2)},\\
&&Z_5=\frac{u(2\piup)^{-d/2}L^{2-d/2}}{(1-d/2)(2-d/2)}\left[(L+Lc)^{2-d/2}-2(L)^{2-d/2}-L_c^{2-d/2}\right].
\end{eqnarray}
Introducing a dimensionless coupling constant $u_0=u(2\piup)^{-d/2}L^{2-d/2}$, and taking into account that $L_c=aL$, we express the partition function in one loop approximation 
as:
\begin{eqnarray}
&&Z^{{\rm pom-pom}}_{f,f}=1-\frac{u_0}{(d-2)(d-4)}\left[4(2+a)^{2-\frac{d}{2}}f_1f_2\right.+4a^{2-\frac{d}{2}}(f_2-1)(f_1-1)\nonumber\\
&&\left.-4(2f_1f_2-f_1-f_2)(a+1)^{2-\frac{d}{2}}+4(f_1^2+f_2^2-f_1-f_2)(2^{1-\frac{d}{2}}-1)\right].\label{Zfinal}
\end{eqnarray}
This expression is used in calculations of all the averaged values that follow below, with averaging defined as:
\begin{eqnarray}
&&\langle (\ldots) \rangle = \frac{1}{{ Z^{\text{pom-pom}}_{f,f}}}\prod_{i=1}^{f}\prod_{j=1}^{f}\,\int\,D\vvec{r}(s)\,\delta(\vvec{r_i}(0)-\vvec{r_0}(0))\delta(\vvec{r_j}(0)-\vvec{r_0}(L))\,{\rm e}^{-H}(\ldots).
\end{eqnarray}
In order to calculate the contribution to the gyration radius in one loop approximation, we have to consider all possible combinations between diagrams in figures \ref{fig:2} and \ref{fig:3}. In general, the expression can be presented as:
\begin{equation}
\langle R^2_{g}\rangle=\langle R^2_{g}\rangle_0\left(1-u_0A(f_1,f_2,a,d)\right) ,   
\end{equation}
with $A(f_1,f_2,a,d)$ being given by the expression:
\begin{eqnarray}
&&A(f_1,f_2,a,d)=-2((f_1+f_2)(3f_1+3f_2-2)+(3(2f_1f_2+f_1+f_2))a+(3(f_1+f_2))a^2+a^3)^{-1}\nonumber\\
&&\times\left(\frac{96(f_1+f_2)(f_1-1)(f_2-1)a^{3-\frac{d}{2}}}{(d-4)(d-2)d(d-6)}\right.+\frac{12f_2f_1(f_1-1)(f_2-1)a^{3-\frac{d}{2}}}{d(d-2)}\nonumber\\
&&+\frac{4a^{4-\frac{d}{2}}(f_2-1)(f_1-1)(f_1+f_2)}{(d-6)d(d-8)(d-2)(d-4)}\times(d^3-18d^2+80d-192)\nonumber\\
&&+\frac{a^{5-\frac{d}{2}}(f_2-1)(f_1-1)(d^2-26d+136)}{(d-10)(d-6)(d-8)(d-4)}-\frac{12(a+1)^{1-\frac{d}{2}}(f_2-1)(f_1-1)2f_1f_2}{d(d-2)}\nonumber\\
&&+\frac{12(a+2)^{1-\frac{d}{2}}(f_2-1)(f_1-1)(4f_1f_2-f_1-f_2)}{d(d-2)}-\frac{12(a+1)^{3-\frac{d}{2}}(2f_1^2f_2^2+f_1^2+f_2^2-f_1-f_2)}{d(d-2)}\nonumber\\
&&-\frac{12(a+1)^{3-\frac{d}{2}}(4d^2-40d+80)f_2f_1}{((d-6)d(d-2)(d-4))}+\frac{12(a+1)^{3-\frac{d}{2}}f_1f_2(3d^2-30d+64)(f_1+f_2)}{(d-6)d(d-2)(d-4)}\nonumber\\
&&-\frac{4(a+1)^{4-\frac{d}{2}}(f_1+f_2-1)(2f_1f_2-f_1-f_2)}{(d-6)d(d-8)(d-2)(d-4)}
(d^3-18d^2+80d-192)\nonumber\\
&&-\frac{(a+1)^{4-\frac{d}{2}}(d^2-26d+136)(2f_1f_2-f_1-f_2)}{(d-10)(d-6)(d-8)(d-4)}
+\frac{(2+a)^{1-\frac{d}{2}}f_1f_2(48d-480)(f_2-1)(f_1-1)}{(d-10)d(d-2)}\nonumber\\
&&-\frac{12(2+a)^{2-\frac{d}{2}}f_1f_2(4f_1f_2-5f_1-5f_2+6)}{d(d-2)}+\frac{12(2+a)^{3-\frac{d}{2}}f_1f_2(f_1f_2-2f_1-2f_2+3)}{d(d-2)}\nonumber\\
&&+\frac{4(2+a)^{4-\frac{d}{2}}f_1f_2(f_1+f_2-2)}{(d-6)d(d-8)(d-2)(d-4)}\times(d^3-18d^2+80d-192)+\frac{(2+a)^{5-\frac{d}{2}}f_1f_2(d^2-26d+136)}{(d-10)(d-6)(d-8)(d-4)}\nonumber\\
&&+\frac{2^{3-\frac{d}{2}}(f_1^2+f_2^2-f_1-f_2)}{(d-10)(d-6)d(d-8)(d-2)(d-4)}((d-10)(d^3-18d^2+8d-192)(a+f_1+f_2)-3840)\nonumber\\
&&+\frac{(3(f_1^2+f_2^2-f_1-f_2))}{(d-10)(d-6)d(d-8)(d-2)(d-4)}(d^4-28d^3+348d^2-2384d+7680)\nonumber\\
&&-\frac{4(d^3-18d^2+128d-576)}{(d-2)d(d-6)(d-4)(d-8)}(f_1f_2(f_1+f_2-2)+(f_1^2+f_2^2-f_1-f_2)a)\nonumber\\
&&-\frac{4(d-12)(d^2-6d+32)}{(d-2)d(d-6)(d-4)(d-8)}\left.(f_2^2(f_2-1)+f_1^2(f_1-1))\right).
\end{eqnarray}

In order to receive the value of the size ratio (\ref{gc}) from the above expression  one can follow one of the two strategies: consider an $\epsilon=4-d$ --- expansion and using the fixed point (\ref{FPP}) or to use a Douglas-Freed approximation \cite{Douglas84}. The first strategy provides only  qualitative values for the ratio and in order to get the quantitative values we need to consider higher orders in $u_0$. The second method allows one to receive values that are comparable with experimental \cite{Douglas84} and numerical \cite{Haydukivska21,Haydukivska22} results.

\ukrainianpart

\title{Особливості розгортання пом-пом полімерів: вплив довжини центрального ланцюжка}

\author{Х. Гайдуківська\refaddr{label1,label2},
	В. Блавацька\refaddr{label1,label3}
}

\addresses{
	\addr{label1} Інститут фізики конденсованих систем Національної академії наук України, \\вул. Свєнціцького, 1, 79011 Львів, Україна
	\addr{label2} Інститут Фізики, Сілезький університет, вул. Першого полку піхоти, 75, 41-500 Хоржув, Польща
	\addr{label3} Діоскурі центр фізики і хімії бактерій,
	Інститут фізичної хімії, Польська академія наук, 01-224 Варшава, Польща
}

\makeukrtitle

\begin{abstract}
	\tolerance=3000%
	Ця стаття є продовженням наших попередніх досліджень пом-пом молекули [K. Haydukivska, O. Kalyuzhnyi, V. Blavatska, 
	and J. Ilnytskyi, J. Mol. Liq. {\bf 328}, 115456 (2021); 
	Condens. Matter Phys. {\bf 25}, 23302 (2022)]. Молекула складається з центрального ланцюжка  з двома центрами галуження на обох кінцях з функціональностями $f_1$ та $f_2$. Тут, основну увагу зосереджено на дослідженні впливу довжини центрального ланцюжка на конфігураційні характеристики складної молекули, такі як характеристики розміру та форми. Ми застосовуємо як схему прямого полімерного перенормування  на основі моделі неперервного ланцюжка, так і альтернативний метод Вея для аналізу низки характеристик розміру і форми пом-пом молекул в слабких розчинниках. Співвідношення форми пом-пом молекули і лінійного полімерного ланцюжка такої ж молекулярної ваги отримано із врахуванням ефекту виключеного об'єму,
	в той час як оцінки для асферичності отримано в Гаусовому наближенні. В той час як для співвідношення розмірів спостерігається монотонна залежність від довжини центрального ланцюжка при різних функціональностях бокових зірок, асферич\-ність демонстує більш нетривіальну поведінку.

\end{abstract}

\end{document}